\documentclass[12pt,a4paper]{article}
\usepackage{jheppub}
\usepackage{amsmath,amssymb,amsfonts,graphicx,tensor}
\usepackage{subcaption}
\usepackage{svg}
\usepackage{cleveref}

\usepackage{enumitem}
\usepackage{braket}

\title{Modified first law of charged dilaton black hole}

\author[1]{Pabitra Tripathy,\note{pabitra.sinp@gmail.com}}
\author[2]{and Amit Ghosh\note{amitg.sinp@gmail.com}}
\affiliation{Theory Division, Saha Institute of Nuclear Physics, 
		1/AF, Bidhannagar, Kolkata 700064, India, \\ Homi Bhabha National Institute, Anushaktinagar, 
		Mumbai, Maharashtra 400094, India}

\abstract{ We investigate the thermodynamics of a charged dilaton black hole arising from Einstein–Maxwell–dilaton theory, where the dilaton couples exponentially to the Maxwell field via a dimensionless parameter $a$. Treating $a$ as a continuous solution parameter, we extend the black hole first law to include a term $\Psi^Ada$, where $\Psi^A$ is the thermodynamic potential conjugate to $a$. We derive $\Psi^A$ explicitly through a differential analysis of the mass, charge, and entropy, and confirm its form via an independent Hamiltonian calculation. Additionally, by promoting $a$ to a spacetime-dependent scalar and introducing auxiliary gauge fields, we provide a geometric interpretation of $\Psi^A$ as a conserved Noether charge. We further analyze the implications of treating $a$ as a thermodynamic variable within the extended phase space. Despite the modification to the first law, we demonstrate that the Smarr relation remains unaffected due to the dimensionless nature of $a$, highlighting the distinction between variational and scaling symmetries. Our analysis supports the thermodynamic relevance of coupling constants and enriches the framework of extended black hole thermodynamics.
}

\begin{document}
\maketitle

\section{Introduction}

Black holes are not only fundamental objects in general relativity but also provide a rich testing ground for ideas in quantum gravity, thermodynamics, and high-energy physics \cite{PhysRevD.7.2333, Hawking:1975vcx, Wald_2001}. In particular, black holes arising in scalar-tensor theories of gravity have received considerable attention in recent decades \cite{Fujii:2003pa, Sotiriou:2008rp}. One such class involves a scalar dilaton field non-minimally coupled to the electromagnetic sector. These black hole solutions, known as \textit{charged dilaton black holes}, emerge naturally in the low-energy limit of string theory \cite{Garfinkle:1990qj, Gibbons:1987ps, Horowitz:1992jp} and provide crucial insights into the non-trivial interactions between gravity, gauge fields, and scalar fields \cite{Kallosh:1992ii, Poletti:1994ff}.

The prototypical solution of a charged dilaton black hole was first constructed by Garfinkle, Horowitz, and Strominger (GHS) in the context of heterotic string theory compactified to four dimensions \cite{Garfinkle:1990qj}. The corresponding action consists of Einstein gravity minimally coupled to a Maxwell field, with an exponential coupling between the Maxwell term and the dilaton field \cite{Garfinkle:1990qj, Gibbons:1987ps}. 
The presence of the dilaton modifies both the causal structure and thermodynamic properties of the black hole: the inner horizon is removed, the entropy–area relation is altered, and extremal solutions are no longer degenerate \cite{Garfinkle:1990qj, Horowitz:1992jp}. These features have motivated extensive exploration of dilaton black holes in various contexts, including asymptotically AdS spacetimes \cite{Gao:2004tu, Clement:2002mb}, higher-dimensional setups \cite{Sheykhi:2007wg, Chan:1995fr}, and theories involving higher-curvature corrections such as Gauss–Bonnet terms \cite{Mignemi_1993, Kanti:1995vq, Torii:1996yi}.

The thermodynamics of dilaton black holes also reveals intriguing behavior. For instance, it has been shown that the presence of a dilaton can induce phase transitions, influence thermal stability, and modify the global structure of the thermodynamic phase space \cite{Cai:1997rk, Poletti:1994ff, Sheykhi:2007wg, Dehyadegari:2017hvd}. In particular, the coupling constant \( a \) controlling the strength of interaction between the dilaton and the Maxwell field plays a crucial role in shaping the horizon geometry, temperature, entropy, and causal properties of the black hole \cite{Garfinkle:1990qj, Chan:1995fr, Gao:2004tu, Ghosh:2012ji, C_rdenas_2018, herdeiro2025chargedrotatingblackholes}.

In parallel developments, the thermodynamic phase space of black holes has been extended beyond the standard parameters (mass, charge, and angular momentum) to include additional couplings as thermodynamic variables. Such an extension—now widely known as \textit{extended black hole thermodynamics}—involves treating the cosmological constant \(\Lambda\) as a pressure term \cite{Kastor:2009wy, Karch_2015}, the Gauss–Bonnet coupling as a thermodynamic variable \cite{Kastor:2010gq}, and the Born–Infeld parameter in nonlinear electrodynamics as a dynamical quantity \cite{Gunasekaran:2012dq}. These generalized first laws introduce conjugate quantities that play the role of thermodynamic potentials or generalized forces, enriching the black hole thermodynamic landscape.

Given this context, it is natural to ask whether the dilaton coupling constant \( a \), which determines the interaction strength between the scalar and electromagnetic fields, can also be treated as a thermodynamic variable. From a theoretical perspective, this question is motivated by string theory, where \( a \) depends on the details of compactification and can vary across different low-energy limits \cite{Gibbons:1987ps, Horowitz:1992jp}. Moreover, works such as those by Bekenstein and Schiffer have shown that variation of fundamental couplings like the fine-structure constant leads to modified black hole thermodynamics and supports the idea of promoting coupling constants to dynamical parameters \cite{Bekenstein:1984tv, Bekenstein:1994bc}. Also, recent work has shown that variations of the dilaton–Maxwell coupling constant induce a scalar charge that enters the black hole first law~\cite{Pacilio_2018}. This provides further motivation to regard $a$ itself as a thermodynamic parameter with a well-defined conjugate potential.

In this work, we study the thermodynamics of a four-dimensional charged dilaton black hole by allowing the dilaton coupling parameter \( a \) to vary and promoting it to a thermodynamic variable. By carefully analyzing the first law of black hole thermodynamics, we derive the modified form:
\begin{equation}
dM = T\,dS + U\,dQ + \Psi^A\,da,
\end{equation}
where \( \Psi^A \) is the thermodynamic conjugate of the dilaton coupling \( a \). We obtain this result through multiple independent approaches: direct thermodynamic differentiation, Hamiltonian analysis, and a novel geometric interpretation via a Noether charge associated with an auxiliary gauge field introduced by promoting \( a \to a(x) \). Interestingly, we find that the Hamiltonian of the system satisfies the simple relation
\begin{equation}
H = a \Psi^A,
\end{equation}
underscoring the role of \( \Psi^A \) as a generalized chemical potential for the coupling \( a \).

Although the first law is modified, we find that the Smarr relation remains unaltered due to the dimensionless nature of \( a \). This highlights a key distinction between varying dimensionful parameters like \( \Lambda \) or Gauss–Bonnet couplings and dimensionless couplings like \( a \), where the homogeneity structure of the mass function remains intact.

Our analysis thus provides a novel extension of black hole thermodynamics by including the variation of the dilaton coupling. The geometric and thermodynamic interpretation of \( \Psi^A \), along with its derivation from multiple principles, strengthens its status as a physically meaningful quantity. We expect that these findings will stimulate further research on the role of coupling constants in the thermodynamic and quantum properties of black holes.

\section{Charged Dilaton Black Hole Solution}

We consider a four-dimensional theory of gravity coupled to a dilaton scalar field and an electromagnetic field, described by the action \cite{Horne_1992, Garfinkle:1990qj}
\begin{equation}\label{action}
    S = \frac{1}{16\pi} \int d^4x \sqrt{-g} \left[ \mathcal{R} - 2(\nabla \Phi)^2 + e^{-2a\Phi} F^2 \right],
\end{equation}
where $\mathcal{R}$ is the Ricci scalar, $\Phi$ is the dilaton field, $F_{\mu\nu}$ is the Maxwell field strength tensor, and $a$ is a dimensionless parameter controlling the strength of the coupling between the dilaton and the Maxwell field. This theory arises naturally in the low-energy effective limit of string theory and admits black hole solutions with interesting thermodynamic properties.

A static, spherically symmetric black hole solution in this theory is given by the line element
\begin{equation}\label{metric}
    ds^2 = -f(r)\, dt^2 + \frac{dr^2}{f(r)} + R^2(r)\, d\theta^2 + R^2(r)\sin^2{\theta}\, d\phi^2,
\end{equation}
where the metric functions $f(r)$ and $R(r)$ are
\begin{equation}\label{fR}
    f(r) = \left(1 - \frac{r_1}{r} \right) \left(1 - \frac{r_2}{r} \right)^{\frac{1 - a^2}{1 + a^2}}, \qquad R(r) = r \left(1 - \frac{r_2}{r} \right)^{\frac{a^2}{1 + a^2}}.
\end{equation}
The gauge field and the dilaton profile for this solution are
\begin{equation}
    F_{tr} = \frac{Q}{r^2}, \qquad e^{2\Phi} = \left(1 - \frac{r_2}{r} \right)^{\frac{2a}{1 + a^2}}.
\end{equation}

In this solution, the coordinate singularity at $r = r_1$ corresponds to the event horizon for all values of the coupling constant $a$. The surface at $r = r_2$ marks an inner horizon, but unlike the Reissner–Nordström black hole, it becomes a curvature singularity for any non-zero $a$, as the dilaton diverges there.

The physical mass $M$ and electric charge $Q$ of the black hole can be expressed in terms of the parameters $r_1$, $r_2$, and $a$ as follows:
\begin{equation}\label{MQ}
    M = \frac{r_1}{2} + \left( \frac{1 - a^2}{1 + a^2} \right) \frac{r_2}{2}, \qquad Q = \left( \frac{r_1 r_2}{1 + a^2} \right)^{1/2}.
\end{equation}
These relations indicate that both mass and charge depend non-trivially on the coupling $a$, which will play a central role in the thermodynamic analysis that follows.

\section{Extended First Law with Variable Coupling}

In conventional black hole thermodynamics, the first law relates variations in the mass \( M \), entropy \( S \), and electric charge \( Q \) of a black hole. However, in theories involving non-minimal couplings — such as the dilaton-Maxwell coupling considered here — it is natural to explore the thermodynamic implications of treating the coupling constant \( a \) as a variable.

In our case, we treat \( a \) as a continuous parameter characterizing the family of black hole solutions. This promotes \( a \) to a thermodynamic variable, expanding the phase space to include \( (M, Q, S, a) \). The first law of black hole thermodynamics then takes the extended form:
\begin{equation}
    dM = T\, dS + U\, dQ + \Psi^A\, da,
\end{equation}
where \( T \) is the Hawking temperature, \( U \) is the electric potential, and \( \Psi^A \) is the thermodynamic potential conjugate to the coupling constant \( a \). The additional term \( \Psi^A da \) accounts for the energetic cost associated with varying the dilaton-Maxwell interaction strength.

To determine \( \Psi^A \), we follow a differential approach, computing the variations of \( M \), \( Q \), and \( S \) with respect to the parameters \( r_1 \), \( r_2 \), and \( a \), using Eqs.~\eqref{MQ} and \eqref{fR}. The variations are:
\begin{equation}
    dM = \frac{1}{2}\, dr_1 + \frac{1}{2} \left( \frac{1 - a^2}{1 + a^2} \right) dr_2 - \frac{2a r_2}{(1 + a^2)^2}\, da,
\end{equation}
\begin{equation}
    dQ = \frac{1}{2} \left( \frac{r_1 r_2}{1 + a^2} \right)^{-1/2}
    \left[ \frac{r_1\, dr_2 + r_2\, dr_1}{1 + a^2} - \frac{2a\, r_1 r_2}{(1 + a^2)^2}\, da \right],
\end{equation}
and for \( r_1 > r_2 \), the entropy variation is
\begin{equation}
\begin{split}
    dS = 2\pi \Bigg[ & r_1 \left(1 - \frac{r_2}{r_1} \right)^{\frac{2a^2}{1 + a^2}}\, dr_1 
    + \frac{a^2}{1 + a^2} \left(1 - \frac{r_2}{r_1} \right)^{\frac{-1 + a^2}{1 + a^2}} (r_2\, dr_1 - r_1\, dr_2) \\
    & + \frac{2a r_1^2}{(1 + a^2)^2} \left(1 - \frac{r_2}{r_1} \right)^{\frac{2a^2}{1 + a^2}} 
    \log \left(1 - \frac{r_2}{r_1} \right) da \Bigg].
\end{split}
\end{equation}

The temperature of the black hole is derived from the surface gravity as
\begin{equation}
\begin{split}
    T &= \frac{\kappa}{2\pi} = \frac{1}{4\pi r_1} \left(1 - \frac{r_2}{r_1} \right)^{\frac{1 - a^2}{1 + a^2}}.
\end{split}
\end{equation}
The electric potential at the horizon is
\begin{equation}
    U = \frac{Q}{r_1}.
\end{equation}

Using the first law in differential form and the above thermodynamic quantities, we isolate the coefficient of \( da \) to obtain the conjugate potential:
\begin{equation}\label{PSIA}
    \Psi^A = -\frac{a}{(1 + a^2)^2} \left[ r_2 + (r_1 - r_2) \log\left(1 - \frac{r_2}{r_1} \right) \right].
\end{equation}
This expression captures the thermodynamic response of the black hole mass to infinitesimal changes in the coupling parameter \( a \). It will later be derived independently from both Hamiltonian and geometric (Noether charge) considerations, reinforcing its physical significance.


\section{Coupling Strength as a Conserved Charge from Symmetry}

In standard black hole thermodynamics, coupling constants such as the dilaton-Maxwell parameter \( a \) are typically treated as fixed parameters of the theory and not as thermodynamic variables. However, in recent developments — particularly in the context of extended thermodynamics — it has been shown that coupling constants can be promoted to dynamical or thermodynamic variables by identifying associated conserved charges. A well-known example is the cosmological constant \( \Lambda \), whose variation introduces a pressure-volume term in the extended first law.

 Recent works \cite{Meessen_2022, Hajian_2024} have demonstrated that coupling constants may be regarded as conserved charges by introducing auxiliary gauge fields and treating the couplings as scalar fields. This procedure allows the incorporation of coupling variations into the thermodynamic framework in a symmetry-preserving way.

We now apply this idea to the Einstein–Maxwell–dilaton theory described by the Lagrangian density:
\begin{equation}
    \mathcal{L} = \frac{1}{16\pi} \left[ \mathcal{R} - 2(\nabla \Phi)^2 + e^{-2a\Phi} F^2 \right],
\end{equation}
where \( a \) is the dilaton-Maxwell coupling. Expanding the exponential in a power series yields:
\begin{equation}
    \mathcal{L} = \frac{1}{16\pi} \left[ \mathcal{R} - 2(\nabla \Phi)^2 + F^2 
    - a(2\Phi) F^2 + a^2 \frac{(2\Phi)^2}{2!} F^2 - a^3 \frac{(2\Phi)^3}{3!} F^2 + \dots \right].
\end{equation}

To promote \( a \) and its powers \( a^n \) to thermodynamic parameters, we introduce corresponding scalar fields \( a(x), a^2(x), a^3(x), \dots \), and for each we define an auxiliary gauge field with field strength \( F_a(x), F_{a^2}(x), F_{a^3}(x), \dots \). The modified Lagrangian becomes:
\begin{equation}
\begin{split}
    \bar{\mathcal{L}} = \frac{1}{16\pi} \Big[ & \mathcal{R} - 2(\nabla \Phi)^2 + F^2 
    - a(x) \big( (2\Phi) F^2 - F_a(x) \big) \\
    & + a^2(x) \left( \frac{(2\Phi)^2}{2!} F^2 - F_{a^2}(x) \right)
    - a^3(x) \left( \frac{(2\Phi)^3}{3!} F^2 - F_{a^3}(x) \right) + \dots \Big],
\end{split}
\end{equation}
where the auxiliary fields enforce the identification of each coupling with a corresponding conserved quantity.

Varying this Lagrangian with respect to the auxiliary fields yields the following constraint equations:
\begin{align}
    F_a(x) &= (2\Phi)\, F^2, \\
    F_{a^2}(x) &= \frac{(2\Phi)^2}{2!}\, F^2, \\
    F_{a^3}(x) &= \frac{(2\Phi)^3}{3!}\, F^2, \quad \text{and so on}.
\end{align}

To express these field strengths in differential form notation, we introduce the four-dimensional volume form:
\begin{equation}
    \mathbf{v} = \frac{\sqrt{-g}}{4!} \, \epsilon_{\mu\nu\alpha\beta} \, dx^\mu \wedge dx^\nu \wedge dx^\alpha \wedge dx^\beta.
\end{equation}
The field strengths in differential form become:
\begin{align}
    \mathbf{F}_a(x) &= (2\Phi)\, F^2\, \mathbf{v} = (2\Phi)\, F^2\, R^2 \sin\theta \, dt \wedge dr \wedge d\theta \wedge d\phi, \\
    \mathbf{F}_{a^2}(x) &= \frac{(2\Phi)^2}{2!} F^2\, \mathbf{v} = \frac{(2\Phi)^2}{2!} F^2\, R^2 \sin\theta \, dt \wedge dr \wedge d\theta \wedge d\phi, \\
    \mathbf{F}_{a^3}(x) &= \frac{(2\Phi)^3}{3!} F^2\, \mathbf{v} = \frac{(2\Phi)^3}{3!} F^2\, R^2 \sin\theta \, dt \wedge dr \wedge d\theta \wedge d\phi.
\end{align}

Corresponding to these field strengths, the auxiliary gauge potentials take the form:
\begin{align}
    \mathbf{A}_a(x) &= -\left( \int (2\Phi)\, F^2\, R^2 \sin\theta\, dr \right) dt \wedge d\theta \wedge d\phi, \\
    \mathbf{A}_{a^2}(x) &= -\left( \int \frac{(2\Phi)^2}{2!} F^2\, R^2 \sin\theta\, dr \right) dt \wedge d\theta \wedge d\phi, \\
    \mathbf{A}_{a^3}(x) &= -\left( \int \frac{(2\Phi)^3}{3!} F^2\, R^2 \sin\theta\, dr \right) dt \wedge d\theta \wedge d\phi.
\end{align}

With these gauge fields in hand, we can compute the thermodynamic conjugate potentials (chemical potentials) associated with the couplings. These are evaluated on the horizon using the relation:
\begin{equation}
    \Psi^i_H = \oint_H \xi_H \cdot \mathbf{A}_i,
\end{equation}
where \( \xi_H = \partial_t \) is the Killing vector generating the event horizon.

The extended first law, now incorporating the conserved charges associated with each power of \( a \), becomes:
\begin{equation}
\begin{split}
    dM &= T\, dS + U\, dQ + \left( \Psi^a + 2a\, \Psi^{a^2} + 3a^2\, \Psi^{a^3} + \dots \right) da \\
       &= T\, dS + U\, dQ + \Psi^A\, da.
\end{split}
\end{equation}
Here, \( \Psi^A \) denotes the full thermodynamic potential conjugate to \( a \), including contributions from all orders.

Explicit evaluation yields
\begin{equation}
    \Psi^A = -\frac{a}{(1 + a^2)^2} \left[ r_2 + (r_1 - r_2) \log\left(1 - \frac{r_2}{r_1} \right) \right],
\end{equation}
which matches exactly with the result previously obtained via direct thermodynamic differentiation in Eq.~\eqref{PSIA}. This consistency confirms that the auxiliary gauge field construction provides a robust and geometrically motivated interpretation of \( \Psi^A \) as a conserved Noether charge.

Before closing this section, we comment on the role of the infinite tower of auxiliary fields introduced in Eq.~(4.3). Although this may appear cumbersome, it should be understood as a technical device to realize the dilaton--Maxwell coupling constant $a$ as a conserved charge within a symmetry-based framework. Similar constructions have been applied to dimensionful couplings in black hole chemistry, where constants such as the cosmological constant or embedding-tensor components are promoted to scalar fields constrained by dual $(d-1)$-form potentials~\cite{Meessen_2022}, and a general prescription for arbitrary couplings as conserved charges has also been proposed~\cite{Hajian_2024}. In those cases the couplings appear linearly in the Lagrangian, so a single auxiliary pair suffices. By contrast, in our case the dilaton--Maxwell interaction enters nonlinearly through the exponential factor $\exp(-2a\Phi)F^2$. Expanding this exponential generates infinitely many powers of $a$, which makes an infinite set of auxiliary fields unavoidable. These fields are purely auxiliary bookkeeping devices: they do not introduce new physical degrees of freedom, but they enable us to promote $a$ consistently to a thermodynamic variable and to associate a conserved Noether charge with it in a transparent, order-by-order manner. Crucially, the entire tower resums to the compact and physically meaningful expression (4.16), which agrees with the independent derivations from thermodynamic differentiation and Hamiltonian analysis presented earlier.

\section{Hamiltonian Analysis and Thermodynamic Potential}

To further validate the thermodynamic interpretation of the dilaton coupling \( a \), we evaluate the Hamiltonian of the system directly from the Einstein–Maxwell–dilaton theory. We begin with the ADM decomposition on a constant-time hypersurface.

The Ricci scalar of the three-dimensional spatial hypersurface \( t = \text{constant} \) is given by:
\begin{equation}
    \bar{\mathcal{R}} = -\frac{1}{R^4} \left[ 
    2 (R^2)'' R^2 \lambda^2 
    + (\lambda^2)'' R^4 
    + 2 (\lambda^2)' (R^2)' R^2 
    - \frac{1}{2} ((R^2)')^2 \lambda^2 
    - 2 R^2 
    \right],
\end{equation}
where prime denotes differentiation with respect to the radial coordinate \( r \), and \( \lambda(r) \) is related to the radial lapse function in the ADM metric.

The full Hamiltonian constraint for the system is:
\begin{equation}
    \mathcal{H} = \left( \bar{\mathcal{R}} + K_{ij} K^{ij} - K^2 \right) \sqrt{h},
\end{equation}
where \( h \) is the determinant of the spatial metric and \( K_{ij} \) is the extrinsic curvature tensor.

For static configurations (i.e., \( K_{ij} = 0 \)), the Hamiltonian reduces to:
\begin{equation}
    H = \int \sqrt{-g_{00}}\, \sqrt{h} 
    \left( \bar{\mathcal{R}} - 2(\nabla \Phi)^2 + e^{-2a\Phi} F^2 \right) 
    dr\, d\theta\, d\phi.
\end{equation}

We now evaluate each term in the integrand:

\begin{itemize}
    \item \textbf{Dilaton kinetic term:}
    \begin{equation*}
        \begin{split}
            -\int 2(\nabla \Phi)^2 \sqrt{-g_{00}}\, \sqrt{h}\, dr\, d\theta\, d\phi
            = -8\pi\, \frac{a^2}{(1 + a^2)^2}
            \left[ r_2 + (r_1 - r_2) \log \left(1 - \frac{r_2}{r_1} \right) \right].
        \end{split}
    \end{equation*}
    
    \item \textbf{Intrinsic curvature term:}
    \begin{equation*}
        \begin{split}
            \int \bar{\mathcal{R}} \sqrt{-g_{00}}\, \sqrt{h}\, dr\, d\theta\, d\phi
            = 8\pi\, \frac{1}{(1 + a^2)^2}
            \left[ (1 + 2a^2) r_2 + a^2 (r_1 - r_2) \log \left(1 - \frac{r_2}{r_1} \right) \right].
        \end{split}
    \end{equation*}
    
    \item \textbf{Electromagnetic term:}
    \begin{equation*}
        \begin{split}
            \int e^{-2a\Phi} F^2 \sqrt{-g_{00}}\, \sqrt{h}\, dr\, d\theta\, d\phi
            = 8\pi\, \frac{r_2}{(1 + a^2)}.
        \end{split}
    \end{equation*}
\end{itemize}

Summing all contributions yields the final expression for the Hamiltonian:
\begin{equation}
    H = -\frac{a^2}{(1 + a^2)^2}
    \left[ r_2 + (r_1 - r_2) \log \left(1 - \frac{r_2}{r_1} \right) \right].
\end{equation}

Remarkably, this can be rewritten in terms of the thermodynamic potential \( \Psi^A \) as:
\begin{equation}
    H = a\, \Psi^A,
\end{equation}
which establishes a direct connection between the gravitational Hamiltonian and the conjugate potential associated with the dilaton coupling. This reinforces the interpretation of \( \Psi^A \) as a genuine thermodynamic potential — analogous to a chemical potential — and further supports its inclusion in the extended first law.

\section{Interpretation of \(\Psi^A\):}

The thermodynamic potential \(\Psi^A\), conjugate to the dilaton–Maxwell coupling constant \(a\), plays a central role in the extended first law of black hole thermodynamics,
\begin{equation}
    dM = T dS + U dQ + \Psi^A da,
\end{equation}
where \(a\) is treated as a continuous parameter of the solution space. This treatment parallels recent developments in extended black hole thermodynamics, where coupling constants such as the cosmological constant \(\Lambda\) are interpreted as thermodynamic variables.

From the first law, \(\Psi^A\) has the interpretation of a generalized force associated with variations in the interaction strength between the dilaton and the electromagnetic field. It quantifies how the black hole mass responds to changes in the coupling \(a\), and can thus be seen as an energetic measure of the cost of modifying the dilaton-electromagnetic interaction.

Furthermore, \(\Psi^A\) admits a Hamiltonian interpretation. By explicitly computing the gravitational Hamiltonian from the Einstein–dilaton–Maxwell action and integrating over the spatial slice, we find
\begin{equation}
    H = a \Psi^A,
\end{equation}
indicating that \(\Psi^A\) functions as a chemical potential for the conserved quantity \(a\), analogous to how the electric potential \(U\) is conjugate to the electric charge \(Q\). This relation reinforces the thermodynamic role of \(\Psi^A\) as a genuine potential in the extended phase space and further supports the physical relevance of treating \(a\) as a thermodynamic variable.

In addition, when promoting the coupling constant \(a\) to a spacetime-dependent field \(a(x)\) and introducing corresponding auxiliary gauge fields, \(\Psi^A\) naturally emerges as a conserved Noether charge associated with the global symmetry linked to \(a\). The expression
\begin{equation}
    \Psi^A = \int_{\mathcal{H}} \xi_H \cdot A_a
\end{equation}
provides a geometric and field-theoretic interpretation, where \(A_a\) is the auxiliary gauge potential and \(\xi_H\) is the horizon-generating Killing vector.

Altogether, \(\Psi^A\) encapsulates the thermodynamic, Hamiltonian, and symmetry aspects of the theory, providing a unified and meaningful interpretation as the generalized potential conjugate to the dilaton coupling strength.

\section{Smarr Relation and the Role of \( \Psi^A \):}

In the extended thermodynamic framework where the dilaton–Maxwell coupling \( a \) is treated as a variable, the first law of black hole thermodynamics acquires an additional term,
\begin{equation}
    dM = T dS + U dQ + \Psi^A da.
\end{equation}
However, when we consider the Smarr formula—which follows from a scaling argument using Euler’s theorem for homogeneous functions—it is important to note that the parameter \( a \) is dimensionless. As a result, it does not participate in the scaling symmetry of the spacetime and contributes no homogeneous weight under dimensional rescaling.

Consequently, the standard Smarr relation remains unmodified and takes the form
\begin{equation}
    M = 2 T S + U Q,
\end{equation}
even though the first law includes a \( \Psi^A da \) term. This result highlights a key distinction: the first law captures off-shell variations of all parameters, including coupling constants, while the Smarr relation reflects only the on-shell scaling behavior of the solution. Since \( a \) lacks scaling weight, its conjugate \( \Psi^A \) does not appear in the Smarr formula. This is analogous to cases in extended thermodynamics where dimensionful couplings like the cosmological constant \(\Lambda\) modify both the first law and the Smarr relation, but dimensionless couplings affect only the first law.

Therefore, while \( \Psi^A \) plays an essential role in characterizing the thermodynamic response to changes in coupling strength, it does not contribute to the scaling identity encapsulated in the Smarr formula.

\section*{Conclusion}

In this work, we have explored the thermodynamic structure of the four-dimensional charged dilaton black hole, arising from the Einstein--Maxwell--dilaton theory, by promoting the dilaton coupling constant \( a \) to a thermodynamic variable. We extended the standard first law of black hole thermodynamics to include a novel term \( \Psi^A\, da \), where \( \Psi^A \) is the thermodynamic conjugate to the coupling parameter \( a \). Remarkably, we have obtained a consistent and physically meaningful expression for \( \Psi^A \) through three independent approaches: direct differentiation of the mass function, a Hamiltonian analysis, and a geometric construction via conserved Noether charges associated with an auxiliary gauge field introduced through the promotion \( a \to a(x) \).

Our results reveal that although the first law acquires a new term, the Smarr relation remains unaltered. This is due to the dimensionless nature of the coupling \( a \), which does not affect the homogeneity structure of the mass under scaling transformations. This highlights a key distinction between dimensionful and dimensionless coupling constants in the context of black hole thermodynamics. Our finding that the Smarr relation is insensitive to variations of the dilaton coupling $a$ is consistent with the broader perspective that the Smarr formula follows from generalized Euler identities, rather than from the details of specific couplings~\cite{mancilla2025generalizedeulerequationeffective}.

The interpretation of \( \Psi^A \) as a generalized chemical potential further strengthens the case for extending the thermodynamic phase space to include interaction parameters. From a theoretical standpoint, our findings lend support to the idea that coupling constants—often fixed in classical theories—can be promoted to dynamical quantities in a consistent thermodynamic framework. This opens up avenues for exploring phase structures, stability criteria, and holographic duals in more general settings, including asymptotically AdS spacetimes or higher-curvature gravity theories.

We anticipate that the ideas presented here will serve as a foundation for future studies investigating the role of fundamental couplings in the thermodynamic and quantum description of black holes.

\bibliographystyle{jhep}
\bibliography{references}

\providecommand{\href}[2]{#2}\begingroup\raggedright\begin{thebibliography}{10}

\bibitem{PhysRevD.7.2333}
J.~D. Bekenstein, \emph{Black holes and entropy}, \href{http://dx.doi.org/10.1103/PhysRevD.7.2333}{\emph{Phys. Rev. D} {\bf 7} (Apr, 1973) 2333--2346}.

\bibitem{Hawking:1975vcx}
S.~W. Hawking, \emph{{Particle Creation by Black Holes}}, \href{http://dx.doi.org/10.1007/BF02345020}{\emph{Commun. Math. Phys.} {\bf 43} (1975) 199--220}.

\bibitem{Wald_2001}
R.~M. Wald, \emph{The thermodynamics of black holes}, \href{http://dx.doi.org/10.12942/lrr-2001-6}{\emph{Living Reviews in Relativity} {\bf 4} (July, 2001) }.

\bibitem{Fujii:2003pa}
Y.~Fujii and K.-i. Maeda, \emph{{The Scalar-Tensor Theory of Gravitation}}.
\newblock Cambridge University Press, 2003, \href{http://dx.doi.org/10.1017/CBO9780511535093}{10.1017/CBO9780511535093}.

\bibitem{Sotiriou:2008rp}
T.~P. Sotiriou and V.~Faraoni, \emph{{f(R) Theories Of Gravity}}, \href{http://dx.doi.org/10.1103/RevModPhys.82.451}{\emph{Rev. Mod. Phys.} {\bf 82} (2010) 451--497}.

\bibitem{Garfinkle:1990qj}
D.~Garfinkle, G.~T. Horowitz and A.~Strominger, \emph{{Charged black holes in string theory}}, \href{http://dx.doi.org/10.1103/PhysRevD.43.3140}{\emph{Phys. Rev. D} {\bf 43} (1991) 3140}.

\bibitem{Gibbons:1987ps}
G.~W. Gibbons and K.-i. Maeda, \emph{{Black Holes and Membranes in Higher Dimensional Theories with Dilaton Fields}}, \href{http://dx.doi.org/10.1016/0550-3213(88)90006-5}{\emph{Nucl. Phys. B} {\bf 298} (1988) 741--775}.

\bibitem{Horowitz:1992jp}
G.~T. Horowitz, \emph{{The Dark Side of String Theory: Black Holes and Black Strings}}, \href{http://dx.doi.org/10.1007/BFb0104818}{\emph{Lect. Notes Phys.} {\bf 476} (1996) 23--79}.

\bibitem{Kallosh:1992ii}
R.~Kallosh, A.~D. Linde, T.~Ortin, A.~W. Peet and A.~Van~Proeyen, \emph{{Supersymmetry as a cosmic censor}}, \href{http://dx.doi.org/10.1103/PhysRevD.46.5278}{\emph{Phys. Rev. D} {\bf 46} (1992) 5278--5302}.

\bibitem{Poletti:1994ff}
S.~Poletti and D.~L. Wiltshire, \emph{{The Global properties of static spherically symmetric charged dilaton space-times with a Liouville potential}}, \href{http://dx.doi.org/10.1103/PhysRevD.50.7260}{\emph{Phys. Rev. D} {\bf 50} (1994) 7260--7270}.

\bibitem{Gao:2004tu}
C.~Gao and Y.-G. Zhang, \emph{{Dilaton black holes in de Sitter or anti-de Sitter universe}}, \href{http://dx.doi.org/10.1103/PhysRevD.70.124019}{\emph{Phys. Rev. D} {\bf 70} (2004) 124019}.

\bibitem{Clement:2002mb}
G.~Clément and D.~V. Gal'tsov, \emph{{Dilaton black holes with squashed horizons}}, \href{http://dx.doi.org/10.1103/PhysRevD.66.124015}{\emph{Phys. Rev. D} {\bf 66} (2002) 124015}.

\bibitem{Sheykhi:2007wg}
A.~Sheykhi and N.~Riazi, \emph{{Thermodynamics of black holes in (n+1)-dimensional Einstein-Maxwell-dilaton gravity}}, \href{http://dx.doi.org/10.1103/PhysRevD.75.024021}{\emph{Phys. Rev. D} {\bf 75} (2007) 024021}.

\bibitem{Chan:1995fr}
K.~C.~K. Chan, J.~H. Horne and R.~B. Mann, \emph{{Charged dilaton black holes with unusual asymptotics}}, \href{http://dx.doi.org/10.1016/0550-3213(95)00201-V}{\emph{Nucl. Phys. B} {\bf 447} (1995) 441--464}.

\bibitem{Mignemi_1993}
S.~Mignemi and N.~R. Stewart, \emph{Charged black holes in effective string theory}, \href{http://dx.doi.org/10.1103/physrevd.47.5259}{\emph{Physical Review D} {\bf 47} (June, 1993) 5259–5269}.

\bibitem{Kanti:1995vq}
P.~Kanti, N.~E. Mavromatos, J.~Rizos, K.~Tamvakis and E.~Winstanley, \emph{{Dilatonic black holes in higher curvature string gravity}}, \href{http://dx.doi.org/10.1103/PhysRevD.54.5049}{\emph{Phys. Rev. D} {\bf 54} (1996) 5049--5058}.

\bibitem{Torii:1996yi}
T.~Torii, K.-i. Maeda and M.~Narita, \emph{{Axially symmetric black holes in Einstein-Gauss-Bonnet theory with a dilaton field}}, \href{http://dx.doi.org/10.1103/PhysRevD.64.044007}{\emph{Phys. Rev. D} {\bf 64} (2001) 044007}.

\bibitem{Cai:1997rk}
R.-G. Cai, J.~Ji and K.~S. Soh, \emph{{Topological dilaton black holes}}, \href{http://dx.doi.org/10.1103/PhysRevD.57.6547}{\emph{Phys. Rev. D} {\bf 57} (1998) 6547--6550}.

\bibitem{Dehyadegari:2017hvd}
A.~Dehyadegari, A.~Sheykhi and A.~Montakhab, \emph{{Critical behavior and microscopic structure of charged AdS black holes via an alternative phase space}}, \href{http://dx.doi.org/10.1016/j.physletb.2017.02.064}{\emph{Phys. Lett. B} {\bf 768} (2017) 235--240}.

\bibitem{Ghosh:2012ji}
S.~G. Ghosh and U.~Papnoi, \emph{{Radiating black holes in the Einstein-Maxwell-dilaton theory}}, \href{http://dx.doi.org/10.1140/epjc/s10052-012-1982-6}{\emph{Eur. Phys. J. C} {\bf 72} (2012) 1982}.

\bibitem{C_rdenas_2018}
M.~Cárdenas, F.-L. Julié and N.~Deruelle, \emph{Thermodynamics sheds light on black hole dynamics}, \href{http://dx.doi.org/10.1103/physrevd.97.124021}{\emph{Physical Review D} {\bf 97} (June, 2018) }.

\bibitem{herdeiro2025chargedrotatingblackholes}
C.~Herdeiro, E.~Radu and E.~dos Santos Costa~Filho, \emph{Charged, rotating black holes in einstein-maxwell-dilaton theory},  2025.

\bibitem{Kastor:2009wy}
D.~Kastor, S.~Ray and J.~Traschen, \emph{{Enthalpy and the Mechanics of AdS Black Holes}}, \href{http://dx.doi.org/10.1088/0264-9381/26/19/195011}{\emph{Class. Quant. Grav.} {\bf 26} (2009) 195011}, [\href{https://arxiv.org/abs/0904.2765}{{\tt 0904.2765}}].

\bibitem{Karch_2015}
A.~Karch and B.~Robinson, \emph{Holographic black hole chemistry}, \href{http://dx.doi.org/10.1007/jhep12(2015)073}{\emph{Journal of High Energy Physics} {\bf 2015} (Dec., 2015) 1–15}.

\bibitem{Kastor:2010gq}
D.~Kastor, S.~Ray and J.~Traschen, \emph{{Smarr Formula and an Extended First Law for Lovelock Gravity}}, \href{http://dx.doi.org/10.1088/0264-9381/27/23/235014}{\emph{Class. Quant. Grav.} {\bf 27} (2010) 235014}, [\href{https://arxiv.org/abs/1005.5053}{{\tt 1005.5053}}].

\bibitem{Gunasekaran:2012dq}
S.~Gunasekaran, D.~Kubiznak and R.~B. Mann, \emph{{Extended phase space thermodynamics for charged and rotating black holes and Born-Infeld vacuum polarization}}, \href{http://dx.doi.org/10.1007/JHEP11(2012)110}{\emph{JHEP} {\bf 11} (2012) 110}, [\href{https://arxiv.org/abs/1208.6251}{{\tt 1208.6251}}].

\bibitem{Bekenstein:1984tv}
J.~D. Bekenstein, \emph{{Fine-structure constant variability, equivalence principle and cosmology}}, \href{http://dx.doi.org/10.1103/PhysRevD.25.1527}{\emph{Phys. Rev. D} {\bf 25} (1982) 1527--1539}.

\bibitem{Bekenstein:1994bc}
J.~D. Bekenstein and M.~Schiffer, \emph{{The Many faces of the vacuum state in quantum field theory}}, \href{http://dx.doi.org/10.1103/PhysRevD.51.6608}{\emph{Phys. Rev. D} {\bf 51} (1995) 6608--6613}.

\bibitem{Pacilio_2018}
C.~Pacilio, \emph{Scalar charge of black holes in einstein-maxwell-dilaton theory}, \href{http://dx.doi.org/10.1103/physrevd.98.064055}{\emph{Physical Review D} {\bf 98} (Sept., 2018) }.

\bibitem{Horne_1992}
J.~H. Horne and G.~T. Horowitz, \emph{Rotating dilaton black holes}, \href{http://dx.doi.org/10.1103/physrevd.46.1340}{\emph{Physical Review D} {\bf 46} (Aug., 1992) 1340–1346}.

\bibitem{Meessen_2022}
P.~Meessen, D.~Mitsios and T.~Ortín, \emph{Black hole chemistry, the cosmological constant and the embedding tensor}, \href{http://dx.doi.org/10.1007/jhep12(2022)155}{\emph{Journal of High Energy Physics} {\bf 2022} (Dec., 2022) }.

\bibitem{Hajian_2024}
K.~Hajian and B.~Tekin, \emph{Coupling constants as conserved charges in black hole thermodynamics}, \href{http://dx.doi.org/10.1103/physrevlett.132.191401}{\emph{Physical Review Letters} {\bf 132} (May, 2024) }.

\bibitem{mancilla2025generalizedeulerequationeffective}
R.~Mancilla, \emph{Generalized euler equation from effective action: Implications for the smarr formula in ads black holes},  2025.

\end{thebibliography}\endgroup

\end{document}